\documentclass{ifacconf}

\usepackage{natbib}
\usepackage{algorithm}
\usepackage{algpseudocode}
\usepackage{graphicx}   
\graphicspath{ {./figures/} }
\usepackage{amsmath}

\usepackage[short]{optidef}
\usepackage{threeparttable}
\usepackage{optidef}
\usepackage{multirow}
\usepackage{threeparttable}
\usepackage{amssymb}
\usepackage{xcolor}
\definecolor{tumblue}{RGB}{0,101,189}
\newcommand{\firstrev}[1]{{\color{blue}{#1}}}
\usepackage{stfloats}
\usepackage{setspace}
\usepackage{tabularx}
\usepackage{booktabs}
\usepackage[T1]{fontenc}
\usepackage[scaled=0.85]{beramono}
\usepackage{listings}
\usepackage{caption}
\usepackage{chemformula}
\usepackage{url}
\usepackage{physics}
\lstdefinestyle{sparqlstyle}{
  language=OCL,
  numbers=left,
  basicstyle=\footnotesize,
  stepnumber=1,
  numbersep=10pt,
  tabsize=2,
  showspaces=false,
  breaklines=true
}
\usepackage[nonumberlist,nopostdot,acronym,toc]{glossaries}
\usepackage{soul}
\usepackage{cases}

\begin{document}
\begin{frontmatter}

\title{Bi-level Model Predictive Control for Energy-aware Integrated Product Pricing and Production Scheduling}

\author[First]{Hongliang Li,} 
\author[First]{Herschel C. Pangborn,} 
\author[First]{Ilya Kovalenko}

\address[First]{The Pennsylvania State University, 
University Park, PA 16802 USA\\
(e-mail: hjl5377@psu.edu, hcpangborn@psu.edu, iqk5135@psu.edu)}

\begin{abstract}  
The manufacturing industry is under growing pressure to enhance sustainability while preserving economic competitiveness.
As a result, manufacturers have been trying to determine how to integrate onsite renewable energy and real-time electricity pricing into manufacturing schedules without compromising profitability.
To address this challenge, we propose a bi-level model predictive control framework that jointly optimizes product prices and production scheduling with explicit consideration of renewable energy availability.
The higher level determines the product price to maximize revenue and renewable energy usage.
The lower level controls production scheduling in runtime to minimize operational costs and respond to the product demand.
Price elasticity is incorporated to model market response, allowing the system to increase demand by lowering the product price during high renewable energy generation.
Results from a lithium-ion battery pack manufacturing system case study demonstrate that our approach enables manufacturers to reduce grid energy costs while increasing profit.
\end{abstract}


\end{frontmatter}


\section{Introduction}
\label{sec:intro}

Manufacturers are facing both economic and environmental pressures that fundamentally challenge operational paradigms~\citep{kovalenko2024harnessing}.
For manufacturers in competitive and cost-sensitive environments, pursuing sustainable goals cannot come at the expense of profitability~\citep{lu2020digital}.
These challenges present a critical trade-off compelling manufacturers to reduce energy costs and mitigate environmental impacts while striving to enhance productivity.
Incorporating renewable energy, such as solar power, into manufacturing systems presents a promising approach for balancing productivity, energy costs, and sustainability objectives.
However, the variability of renewable energy poses challenges for production scheduling.
Manufacturers need to coordinate production operations not only in response to product demand, but also in alignment with renewable energy availability.

One promising approach to align production schedules with renewable energy availability is to strategically lower product prices when renewable energy supply is abundant.
This approach creates an economic incentive for customers to increase their demand during periods of high renewable generation when energy is cleaner and less costly.
The underlying rationale is based on the concept of price elasticity, which characterizes how product demand responds to price changes.
Specifically, in make-to-stock manufacturing systems, lowering prices typically stimulates higher demand~\citep{fibich2005dynamics}.
However, incorporating such dynamic pricing strategies into production and energy management introduces the need for joint optimization, adding complexity to manufacturing system scheduling.
Moreover, product pricing decisions typically operate on longer time horizons than operational production scheduling.
Therefore, effectively coupling pricing with scheduling requires a control framework across different time scales.

Bi-level Model Predictive Control (MPC) offers a promising framework for addressing scheduling complexities, as this approach explicitly considers the hierarchical decision structure where strategic and operational decisions are optimized at different levels~\citep{10886496}. At the higher level, strategic pricing decisions can be optimized to maximize profit.
At the lower level, an energy-aware production schedule can be computed in response to the pricing strategies and resulting product demand. 
This hierarchy reflects real-world manufacturing decision-making structures where different organizational levels balance conflicting objectives across varying time scales and priorities~\citep{VANDEBERG2024108726}.
Despite potential benefits, research that jointly integrates product pricing, production scheduling, and renewable energy management is limited.
Existing work focuses on pairwise combinations, such as pricing with scheduling~\citep{CHEN201813} or scheduling with energy management~\citep{li2023system}.

This paper proposes a bi-level MPC framework, shown in Fig.~\ref{fig:framework}, that jointly optimizes product pricing and production scheduling while explicitly accounting for renewable energy integration.
The proposed framework will enable manufacturers to coordinate product pricing and production scheduling to meet economic and environmental goals.
The key contributions are: (1) a network-based manufacturing system model that captures manufacturing dynamics and energy consumption, (2) an approximated gradient-based solution approach for bi-level MPC that handles the Mixed-Integer Quadratic Program (MIQP) of the lower-level scheduling problem, and (3) explicit integration of renewable energy into the decision-making process for production scheduling.
The remainder of the paper is organized as follows. Section~\ref{sec:problem} presents the problem statement and assumptions. Section~\ref{sec:model} describes the system model. Section~\ref{sec:methods} details the bi-level MPC formulation. Section~\ref{sec:case} presents case study results. Section~\ref{sec:conclusion} concludes with future work directions.

\section{Problem Statement}
\label{sec:problem}

\begin{figure*}
    \centering
    \includegraphics[width=0.95\linewidth]{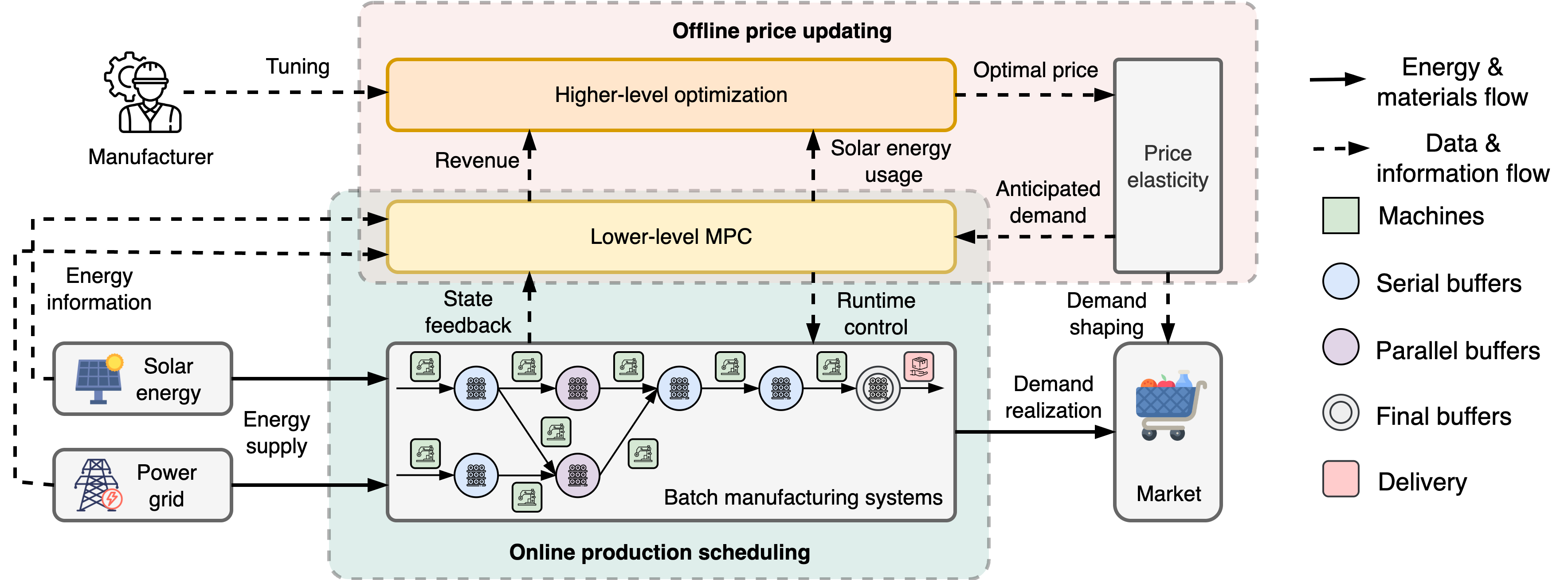}
    \caption{Bi-level MPC framework for integrated product pricing and production scheduling with onsite solar energy in a make-to-stock batch manufacturing environment.}
    \label{fig:framework}
\end{figure*}

We consider a make-to-stock batch manufacturing system as illustrated in Fig.~\ref{fig:framework}. The manufacturer determines the selling price at the beginning of each daily planning horizon, which influences total demand through price elasticity~\citep{ha1997optimal}. Production activities are then scheduled hourly to fulfill this demand. We formalize this as a joint pricing and scheduling optimization problem with three goals: (1) determine the optimal product pricing to shape daily demand, (2) schedule machine operations for minimal cost and demand fulfillment, and (3) efficiently integrate renewable energy to reduce grid dependency.

To simplify the modeling and focus on key decision interactions, we adopt the following assumptions. First, there are no constraints on raw material supply. This assumption allows us to isolate the impact of energy costs and demand variations on scheduling decisions without the added complexity of supply chain disruptions. Second, all energy requirements are met using electricity sourced from the power grid and solar photovoltaic (PV)  panels. This situation represents a typical industrial setup where renewable energy supplements grid power. Finally, the operational cost of solar panels is not considered while electricity drawn from the grid is priced based on Real-Time Pricing (RTP). In practice, renewable sources have negligible marginal costs once the infrastructure is installed.

\section{System Model}
\label{sec:model}

\subsection{Manufacturing Network Model}

The discrete-time state-space representation of the networked batch manufacturing system dynamics is:
\begin{equation}
    x(k+1) = Ax(k) + Bu(k) + Wd(k)
\end{equation}
where $x(k) \in \mathbb{R}^{n_x}$ is the vector of buffer levels, $u(k) \in \mathbb{R}^{n_u}$ is the vector of machine processing rates, and $d(k) \in \mathbb{R}^{n_p}$ is the vector of final product outflows at time step $k$. 
$n_x,n_u,n_d$ are the total number of buffers, machines, and products.
We index buffers, machines, and products by $i$, $j$, and $p$, respectively.
The state transition matrix $A\in\mathbb{R}^{n_x \times n_x}$ is the identity matrix.
The input matrix $B \in \mathbb{R}^{n_x \times n_u}$ captures the network topology:
\begin{equation}
    B_{ij} = 
    \begin{cases}
        1, & \text{if input $j$ flows into buffer $i$} \\
        -1, & \text{if input $j$ flows out of buffer $i$} \\
        0, & \text{otherwise}
    \end{cases}
\end{equation}
The matrix $W \in \mathbb{R}^{n_x \times n_p}$ maps final product outflows from buffer states:
\begin{equation}
    W_{ip} = 
    \begin{cases}
        -1, & \text{if product $p$ flows out of buffer $i$} \\
        0, & \text{otherwise}
    \end{cases}
\end{equation}
The matrix $B_o$ tracks material outflows, defined as $B_{o,ij} = -1$ if $B_{ij} < 0$ and $B_{o,ij} = 0$ otherwise.

\subsection{Energy Consumption Model}
We define energy consumption as $\epsilon_j$ (kWh/unit) for process $j$. The energy usage of time step $k$ is then given by:
\begin{equation}
    \label{eq:machine-energy-usage}
    E(k) = \sum_{j=1}^{n_u} \epsilon_j \cdot u_j(k) \cdot \Delta t
\end{equation}
where $\Delta t$ is the time step interval.
The total energy is from both renewable energy $E_{r}(k)$ and grid energy $E_{g}(k)$:
\begin{equation}
\label{eq:machine-energy-usage_relation}   
        E(k) = E_{g}(k) + E_{r}(k),\;
        E_{r}(k) \leq E_{ar}(k)
\end{equation}
where $E_{ar}(k)$ is the available renewable energy.
The operational cost of renewable energy is not considered, as its marginal cost becomes negligible once the infrastructure is installed, making it insignificant compared to the grid energy cost.
Let $\rho_e(k)$, for $k=0,\dots,N-1$, denote the RTP over a prediction horizon of length $N$. The total energy cost is given by:
\begin{equation}
    C_{E} =\sum_{k=0}^{N-1} \rho_e(k)E_{g}(k)
\end{equation}

\subsection{Price Elasticity Model}
We use an affine price elasticity function to model the correlation between product price and customer demand in a make-to-stock supply chain:
\begin{equation}
\label{eq:price-elasticity}
    \gamma = a - b \cdot p
\end{equation}
where $a$ is the base demand potential, $b$ is the price sensitivity coefficient, $\gamma$ is the potential demand, and $p$ is the product price.
More details about the theoretical foundations and empirical applications of affine price elasticity functions can be found in~\cite{mankiw2021principles}.

\subsection{System Operation Constraints}
\subsubsection{Buffer and Machine Capacity Constraint:}
Buffers and machines are constrained by their capacity:
\begin{equation}
    x_{\min} \leq x(k) \leq x_{\max},\; u_{\min} \leq u(k) \leq u_{\max}
\end{equation}
\noindent where
$x_{\min}$, $x_{\max}$, $u_{\min}$, and $u_{\max}$ are the minimum and maximum buffer and processing capacities, respectively.

\subsubsection{Machine On-off Constraint:}
We model the machine's operational state as $\delta(k) \in \{0,1\}$, e.g., $\delta(k)=1$ indicates that the machine is \textit{on} at step $k$. 
To track machine startups, we introduce a binary variable $\delta_{on}(k) \in \{0,1\}$ which is 1 when the machine transitions from the \textit{off} to the \textit{on} state and subject to the following constraints:
\begin{equation}
    \label{eq:on-off-1}
    \delta_{on}(k) \leq \delta(k)
\end{equation}
\begin{equation}
    \label{eq:on-off-2}
    \delta(k) - \delta(k-1) \leq \delta_{on}(k) \leq \mathbf{1} - \delta(k-1)
\end{equation}
These constraints enforce $\delta_{on}(k)=1$ when the machine turns \textit{on}. 
To reduce wear, the machine must run for a minimum duration $\varrho$ after activation:
\begin{equation}
    \label{eq:min-run}
    \delta(k+t) \geq \delta_{on}(k), \quad \forall t \in \{0,1,\ldots,\varrho-1\}
\end{equation}
Constraint~\eqref{eq:min-run} ensures that if the startup occurs at step $k$, then the machine remains \textit{on} for steps $k$ through $k+\varrho-1$.

\subsection{Production Requirement Constraints}
We formulate dynamic production requirement constraints that progressively tighten production deviation from the target as deadlines approach.
Let $N$ and $\lambda$ denote the MPC prediction horizon and completed production at the initial time step of the MPC.
Let $\pi(k)$ denote the predicted cumulative production up to step $k$:
\begin{equation}
    \pi(k) =\sum_{t=0}^{k} d(t), \; k \in \{0, \dots, N-1\}
\end{equation}
We introduce non-negative slack variables $s(k)$ to allow temporary, penalized deviations from the end-of-horizon production targets $\gamma$:
\begin{equation}
    \pi(k) \geq \gamma - \lambda - s(k)
\end{equation}
To enforce increasingly tight adherence to production targets as the deadline nears, the slack variable is constrained:
\begin{equation}
    0 \leq s(k) \leq \alpha(k) \cdot \gamma
\end{equation}
where $\alpha(k)$ is the allowable deviation as a percentage of the production target $\gamma$.
The time-varying allowable slack percentage is calculated using a heuristic metric:
\begin{equation}
    \alpha(k) = \tau \cdot (1 - \eta(k) \cdot (1 - \xi))
\end{equation}
where $\tau$ is the base production tolerance and $\xi \in[0,1]$ is the tightening factor. 
Let $h$ denote the absolute time step corresponding to the start of the production and $H$ denote the overall production period.
The progress metric $\eta(k)$ depends on the absolute time step:
\begin{equation}
    \eta(k) = \frac{h+k}{H}
\end{equation}
This heuristic provides the controller more flexibility during earlier periods, i.e., when $\eta(k)$ is small, $\alpha(k) \approx \tau$.
On the other hand, the controller ensures that production converges towards the required targets as the final deadline $H$ approaches, i.e., when $\eta(k) \to 1$, $\alpha(k) \to \tau \cdot \xi$.

\section{Bi-Level MPC Formulation}
\label{sec:methods}

Fig.~\ref{fig:framework} shows the bi-level MPC framework structure.
The higher-level optimization is executed once per day in open-loop using day-ahead solar energy availability and RTP data.
The higher level solves a bi-level optimization problem where the Lower-level MPC (L-MPC) is embedded as an inner problem to evaluate the production revenue, renewable energy usage, and feasibility of candidate pricing strategies.
This offline process yields the daily product price $p$ and its corresponding anticipated daily demand $\gamma$.
The L-MPC operates online, solving a shrinking-horizon MIQP at each control step using updated buffer level and machine operation measurements to generate optimal production control actions that fulfill $\gamma$.
This hierarchical structure operates across different time scales, with daily pricing decisions driving hourly production scheduling.

\subsection{Bi-level Problem Formulation}
\subsubsection{Lower-level MPC:}
The L-MPC uses a shrinking horizon strategy to generate production schedules.
The L-MPC solves the following MIQP to meet the product demand:
\begin{subequations}
\label{eq:lower-level}
    \begin{align}
        \min_{u, d, \delta, \delta_{on}, s} &\quad \sum_{k=0}^{N-1} \Big[ \|x(k) - x_g\|_{Q_g}^2 + \omega_e \rho_e(k) E_g(k)\\
        &\quad  + c_{on}{^T}\delta_{on}(k) - \omega_r E_r(k) + \omega_s |s(k)|\Big] \notag \\
        &\quad  + \|x(N) - x_t\|_{Q_t}^2 \label{eq:lower} \\
        \text{s.t.} 
        &\quad \forall k \in \{0,1,\ldots,N-1\}\\
        &\quad x(k+1) = Ax(k) + Bu(k) + Wd(k) \label{eq:dynamics} \\
        &\quad Ax(k) + B_ou(k) + Wd(k)\geq 0 \label{eq:material1} \\
        &\quad d(k) \geq 0 \label{eq:material2} \\
        &\quad x_{\min} \leq x(k) \leq x_{\max} \label{eq:state_bounds} \\
        &\quad \delta(k)u_{\min} \leq u(k) \leq \delta(k)u_{\max} \label{eq:input_bounds} \\
        &\quad \delta(k) - \delta(k-1) \leq \delta_{on}(k) \leq \delta(k) \label{eq:startup1} \\
        &\quad \delta_{on}(k) \leq \mathbf{1} - \delta(k-1) \label{eq:startup2} \\
        &\quad \delta(k+t) \geq \delta_{on}(k), \forall t \in \{0,\ldots,\varrho-1\} \label{eq:min_run} \\
        &\quad \pi(k) \geq \gamma - \lambda - s(k) \label{eq:prod1} \\
        &\quad 0 \leq s(k) \leq \alpha(k)\gamma \label{eq:prod2}
        \end{align}
\end{subequations}
\noindent
The daily production demand $\gamma$ is determined from the price elasticity~\eqref{eq:price-elasticity}.
The objective function is developed with six terms: buffer holding costs (weighted by $Q_g$), grid energy costs (weighted by $\omega_e$), machine startup costs, renewable energy promotion (weighted by $\omega_r$), slack penalties (weighted by $\omega_s$), and terminal buffer regulation (weighted by $Q_t$).
The parameters $x_g$ and $x_t$ represent desired buffer levels during and at the end of the L-MPC prediction horizon, respectively.
The parameter $c_{on}$ is the machine startup cost.

\subsubsection{Higher-level Optimization:}
The higher-level optimization determines the product price based on the amount of renewable energy used and revenue during production. The higher-level problem is formulated as:
\begin{subequations}
\label{eq:higher-level}
    \begin{align}
        \min_{p} &\quad \sum_{k=0}^{H-1} \Big[ - p(k)^T d^*(k) - E_{r}(k) \Big]
        \label{eq:higher} \\
        \text{s.t.} 
        &\quad p_{min} \leq p(k) \leq p_{max} \\
        &\quad [u^*(k),d^*(k)] \; \text{is optimal for L-MPC}
        \end{align}
\end{subequations}
\noindent Note that $E_{r}(k)$ is calculated using $u^*(k)$ by~\eqref{eq:machine-energy-usage}-\eqref{eq:machine-energy-usage_relation}.
$H$ is the production horizon.
The first term in the objective function maximizes revenue and the second term encourages the use of renewable energy.

\begin{algorithm}[t]
\caption{Component-wise Approximate Gradient Method for Bi-level MPC}
\begin{algorithmic}[1]
\Require Initial price $p_0$, tolerance $\epsilon$, maximum iteration $K_{max}$, parameters for problems~\eqref{eq:lower-level} and~\eqref{eq:higher-level}
\Ensure Price $p^*$, and schedule $u^*$, $d^*$
\State Initialize $k \gets 0$, $p_k \gets p_0$, Converged $\gets$ False
\While{$k\leq K_{max}$ and \textbf{not} converged}
    \State Calculate demand by $\gamma_k = a-bp_k$ 
    \State Solve L-MPC problem~\eqref{eq:lower-level}
    \State Obtain $u^*_k, d^*_k$, $\iota$
    \State Update price by
    \begin{equation*}
        p_{k+1} = \Pi_{\mathcal{P}}[p_k -\kappa_{u}\mathcal{G}_d(p_k)
            -\kappa_{d}\mathcal{G}_d(p_k)]
    \end{equation*}
    \If{$|p(k+1)-p(k)|\leq \epsilon$}
        \State Converged $ \gets$ True
    \EndIf
    \State $k \gets k+1$
\EndWhile
\State $p^* \gets p_{k}$
\State $u^*,d^* \gets$ Solve L-MPC problem~\eqref{eq:lower-level} with $p^*$\\
\Return $p^*, u^*, d^*$
\end{algorithmic}
\label{alg:pgm}
\end{algorithm}

\subsection{Component-wise Approximate Gradient Method}
The bi-level MPC is challenging to solve due to the inherent nested optimizations.
Existing approaches, such as KKT reformulation and standard gradient-based methods, are intractable for our problem due to the mixed-integer variables and nonconvexities in the lower-level MIQP problem.
We propose a heuristic solution approach with approximated gradient information, inspired by~\citep{chen2021tighter,10886496}.
The proposed algorithm iteratively uses component-wise approximate gradient information to solve the bi-level problem until convergence.

We define the higher-level and lower-level optimization as the implicit function of the decision variables:
\begin{equation}
    \Phi^* = \underset{\mathcal{P}}{\text{argmin}}\, F(p, u^*(p), d^*(p))
\end{equation}
where $\mathcal{P}$ is the feasible set of prices, and $u^*(p)$ and $d^*(p)$ are the optimal production schedule and product delivery at price $p$, respectively, and
\begin{equation}
    \Psi^* = \underset{\Omega}{\text{argmin}}\, L(f(p),u,x,d,\delta_{on})
\end{equation}
where $\Omega$ is the feasible set of production schedules, and $f(p)$ is the price-dependent demand function.
The standard projected gradient method involves iterative updates:
\begin{equation}
    p_{k+1} = \Pi_{\mathcal{P}}[p_k - \alpha_k \nabla_pF(p_k)]
\end{equation}
The hyper-gradient $\nabla_pF$ is given by~\citep{chen2021tighter}:
\begin{equation}
    \nabla_pF = \frac{\partial F}{\partial p} + \frac{\partial F}{\partial u}\frac{\partial u}{\partial p} + \frac{\partial F}{\partial d}\frac{\partial d}{\partial p}
\end{equation}
In our context, computing $\partial u/\partial p$ and $\partial d/\partial p$ exactly is challenging due to the MIQP structure of the lower level.
We propose Algorithm~\ref{alg:pgm}, which employs a decomposed approximation of the gradient direction:
\begin{equation}
    \label{eq:grad_approx}
    \nabla_pF(p_k) \approx \kappa_{u}\mathcal{G}_u(p_k) + \kappa_{d}\mathcal{G}_d(p_k)
\end{equation}
where the gradient components $\mathcal{G}_u$ and $\mathcal{G}_d$ are approximated gradient directions related to the two parts of the objective $F$, namely the revenue and renewable energy usage.
$\kappa_u, \kappa_d$ are non-negative weights with $\kappa_u+\kappa_d=1$.

\subsubsection{Renewable Energy Usage Gradient Component:}
The renewable energy gradient component quantifies how price changes should affect renewable energy usage:
\begin{equation}
\mathcal{G}_u(p_k) = (1-\frac{1}{\iota^*}\iota(p_k))(p_{\max} - p_{\min})
\end{equation}
where $\iota \in [0,1]$ represents the fraction of energy from renewable sources. 
This formulation creates a sign-changing gradient that encourages price increases when renewable utilization is below the desired percentage $\iota^*$ and price decreases when renewable utilization is high.

\subsubsection{Revenue Gradient Component:}
Revenue is calculated by multiplying the realized product demand $d_k$ and price $p_k$.
The L-MPC controls the production schedule to meet the anticipated demand $\gamma_k$, hence we approximate the revenue by multiplying $\gamma_k$ and $p_k$.
Based on the price elasticity model $\gamma_k = a-bp_k$, the approximated revenue gradient is:
\begin{equation}
    \mathcal{G}_d(p_k) = \frac{\mathrm{d}(p_k \cdot d_k)}{\mathrm{d}p_k}=\frac{\mathrm{d}(p_k(a - bp_k))}{\mathrm{d}p_k} = a - 2bp_k
\end{equation}
This gradient points toward the revenue-optimal price.
\subsubsection{Combined Update Rule and Projection:}
The algorithm combines these gradient components based on~\eqref{eq:grad_approx}:
\begin{equation}
p_{k+1} = \Pi_{\mathcal{P}}[p_k -\kappa_{u}\mathcal{G}_u(p_k)- \kappa_{d}\mathcal{G}_d(p_k)]
\end{equation}
The projection operator $\Pi_{\mathcal{P}}$ ensures that the updated price remains within the feasible range:
\begin{equation}
\Pi_{\mathcal{P}}[p_k] = \max[p_{\min}, \min(p_k, p_{\max})]
\end{equation}
This approach transforms the bi-level optimization into a sequence of tractable subproblems solved using the proposed component-wise approximated gradients.
\section{Case Study}
\label{sec:case}

\subsection{Case Study Setup}
\begin{figure}[t]
    \centering
    \includegraphics[width=1\linewidth]{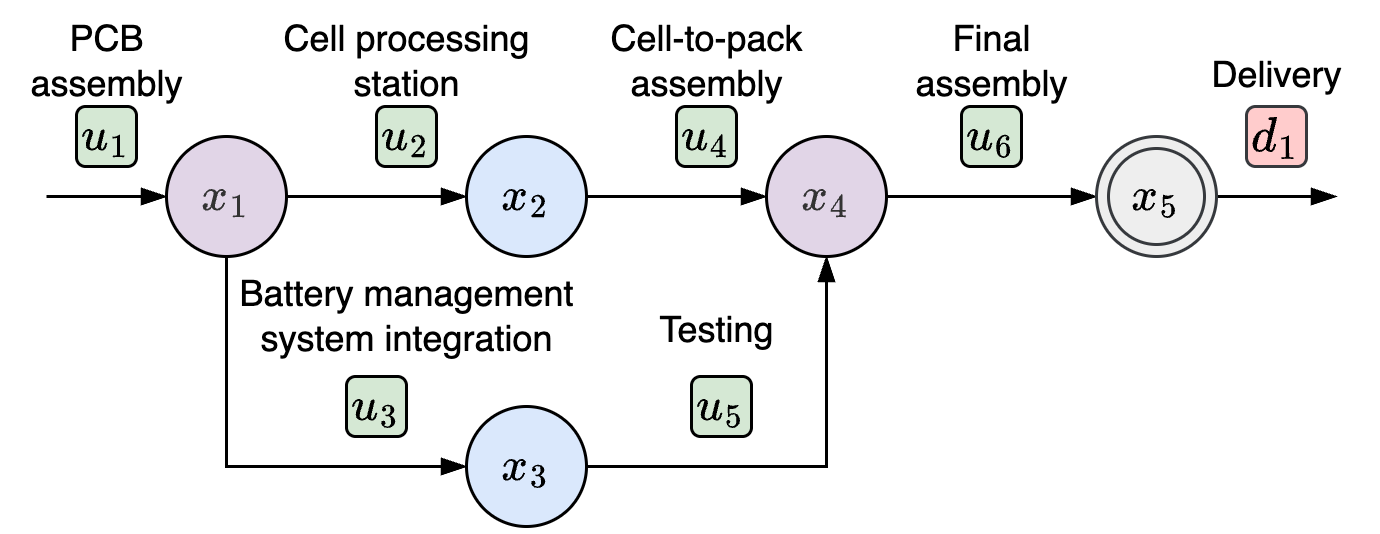}
    \caption{Lithium-ion battery pack manufacturing system considered in the case study.
    }
    \label{fig:case_sys}
\end{figure}

We consider a lithium-ion battery pack manufacturing system as shown in Fig.~\ref{fig:case_sys}. The system parameters are designed based on~\cite{heimes2018lithium}.
The maximum machine processing rates are $u_{\max,1} = 10$ units/hour for Printed Circuit Board (PCB) assembly and $u_{\max,i} = 8$ units/hour for subsequent operations ($i=2,\dots,6$). Buffer capacities are $x_{\max,i} = 20$ units for intermediate buffers ($i=1,\dots,4$) and $x_{\max,5} = 40$ units for finished product inventory. The L-MPC weights are $Q_g = \text{diag}(0.01, 0.01, 0.01, 0.02, 0.2)$ and $Q_t = \text{diag}(0.05, 0.05, 0.05, 0.1, 1.0)$.
Machine startup costs are $c_{on} = [2.0\; 3.0\; 2.5\; 2.8\; 3.5\; 3.0]^T$ USD, power consumption is $\epsilon_j = [3.5\; 8.2\; 6.7\; 4.3\; 5.8\; 9.5]^T$ kW/unit, and processing times are $\varrho = [3\; 4\; 3\; 2\; 4\; 3]^T$ hours. All desired buffer levels are set to zero.
Additional parameters are listed in Table~\ref{tab:sim_params}.

\firstrev{
\begin{table}
\centering
\caption{Simulation Parameters}
\label{tab:sim_params}
\begin{tabularx}{\columnwidth}{@{} l >{\raggedright\arraybackslash}X l @{}} 
\toprule 
\textbf{Symbol} & \textbf{Parameter} & \textbf{Value} \\
\midrule 
$N$ & Prediction horizon (hour) & 24 \\
$a$ & Base demand potential (unit) & 120 \\
$b$ & Price sensitivity (unit/USD) & 0.8 \\
$p_{min}$ & Minimum price (USD) & 70\\
$p_{max}$ & Maximum price (USD) & 120\\
$\xi$ & Tolerance tightening factor & 0.5 \\
$\tau$ & Production tolerance & 0.05 \\
$\iota_*$ & Desired percentage of solar energy & 0.5 \\
$\omega_e$ & Grid energy weight & 10.0 \\
$\omega_r$ & Renewable energy weight & 5 \\
$\omega_s$ & Slack penalty & 5000 \\
$\kappa_{d}$ & Revenue gradient weight & 0.6 \\
$\kappa_{u}$ & Renewable gradient weight  & 0.4 \\
$K_{max}$ & Maximum iterations  & 30 \\
\bottomrule 
\end{tabularx}
\vspace{1ex}
\raggedright 
\end{table}
}

Two scenarios are compared in the case study: (1) a baseline case using only grid electricity, and (2) a solar-integrated case incorporating onsite PV generation.
The simulation covers 5 days using real energy data from the Chicago area, Illinois, USA, during May 2020.
RTP data is acquired from the PJM market operator~\citep{PJM_website}.
Solar energy availability is calculated using daily solar radiation data from OpenEI and solar calculator, assuming a 150 kW direct current system (833 $m^2$ panel area with 18\% efficiency)~\citep{OpenEI_website}.
The simulation framework is implemented in MATLAB, with the MIQP problems for the L-MPC solved using the Gurobi optimizer~\citep{gurobi}.

\subsection{Results and Discussions}

\begin{figure}[t]
    \centering
    \includegraphics[width=0.95\linewidth]{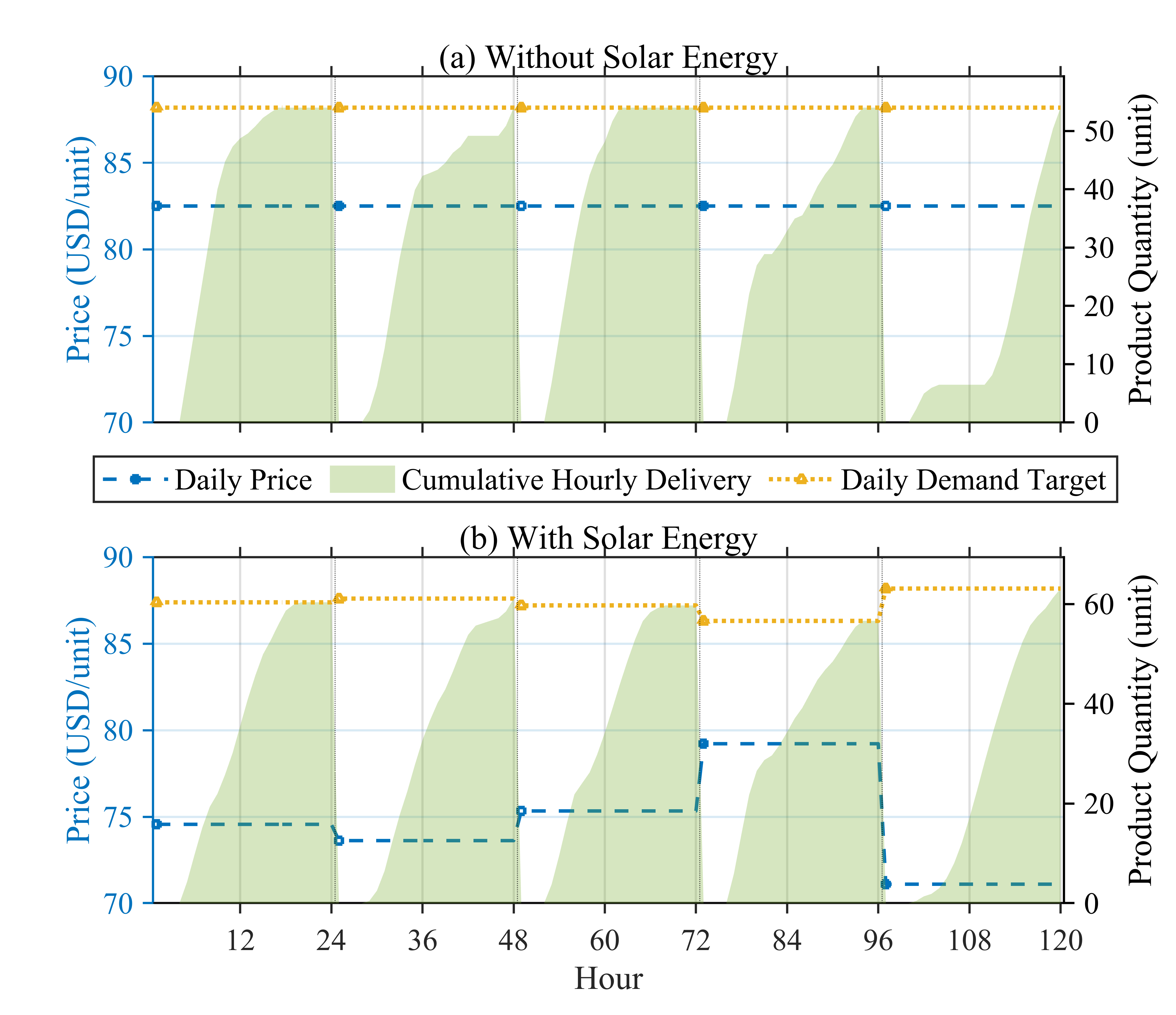}
    \caption{Daily price, demand, and cumulative hourly product delivery over a 5-day horizon.}
    \label{fig:price-production}
\end{figure}

Fig.~\ref{fig:price-production} illustrates the daily pricing decisions, daily demand, and cumulative production over the 5-day horizon for both scenarios.
In the baseline case, the price remains constant at 82.50 USD/unit.
In contrast, the solar-integrated scenario adopts a dynamic pricing strategy, lowering prices to as low as 71.11 USD during periods of high solar availability, with an average price of 74.77 USD representing a 9.4\% reduction compared to the baseline.
The shaded area shows the cumulative hourly delivery.
Results show that daily demand targets are consistently satisfied in both scenarios.
The lower prices in the solar-integrated case stimulated additional demand, resulting in an 11.5\% increase in total production.
This increase is achieved without sacrificing profitability, as the system leverages reduced energy costs from solar power to offer competitive pricing while improving overall economic performance.
\begin{figure}[t]
    \centering
    \includegraphics[width=1\linewidth]{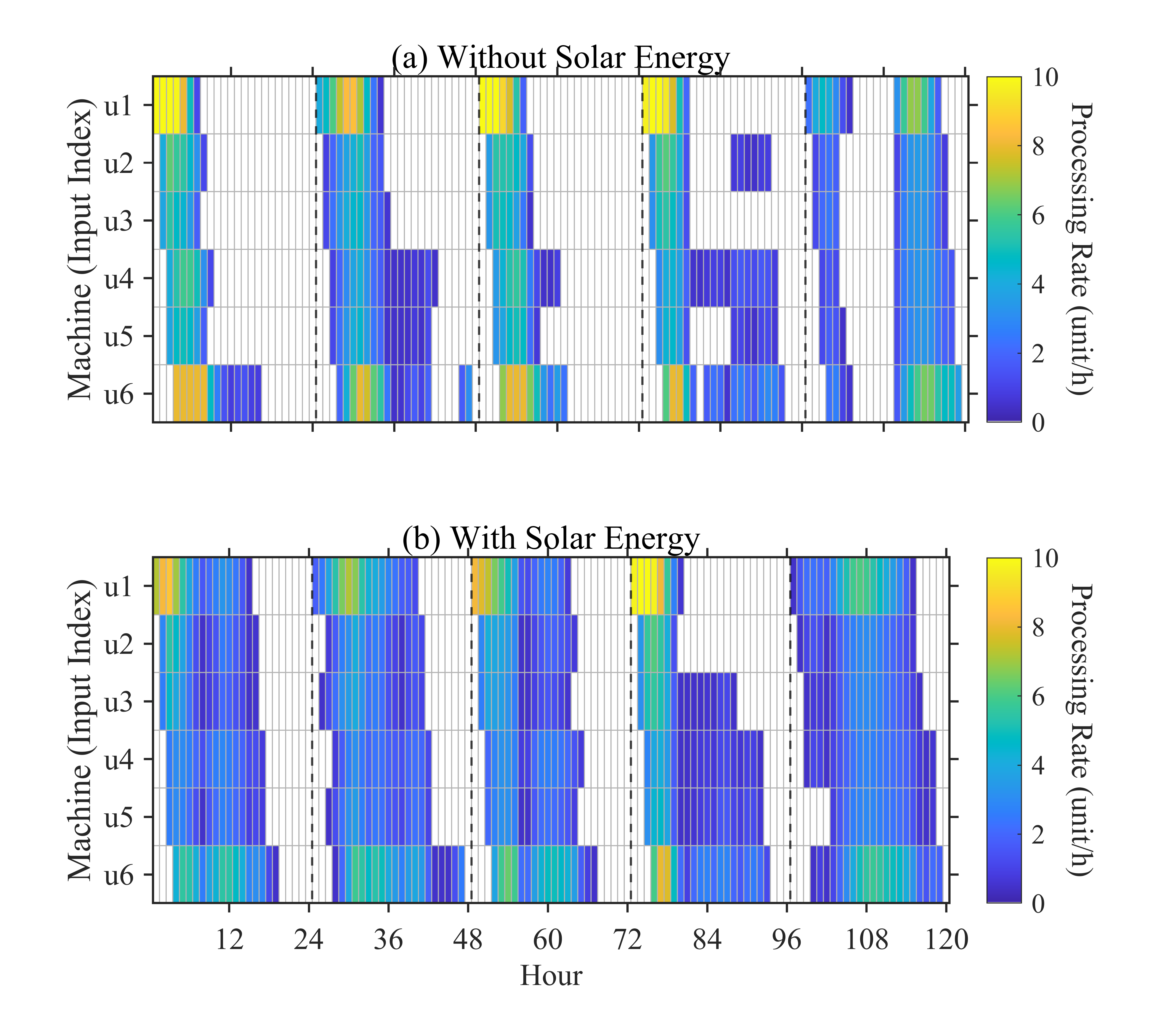}
    \caption{Machine operation schedules (machine on-off and processing rate) over a 5-day horizon.}
    \label{fig:machine-operations}
    \vspace{-2pt}
\end{figure}

\begin{figure}[t]
    \centering
    \includegraphics[width=1\linewidth]{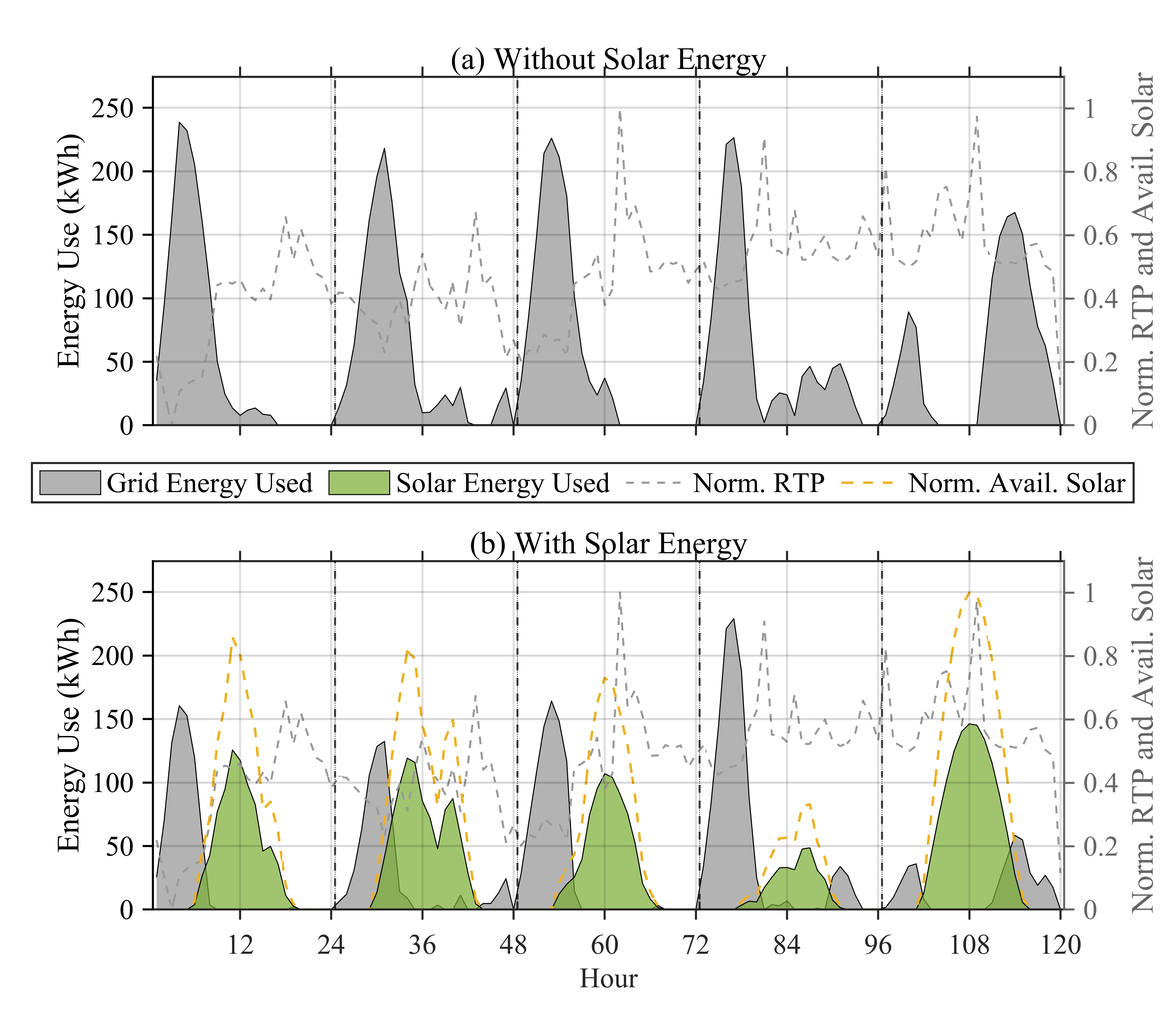}
    \caption{Grid and solar energy consumption over a 5-day horizon. Normalized RTP and solar energy availability are shown as dashed lines.
    }
    \label{fig:energy-usage}
\end{figure}

Fig.~\ref{fig:machine-operations} illustrates the machine schedules for both scenarios, with color indicating the processing rate.
The baseline case exhibits more frequent startups and shutdowns in response to RTP, resulting in higher startup costs.
In contrast, the solar-integrated system demonstrates more consistent production, improving both energy efficiency and system-level cost-effectiveness.
Fig.~\ref{fig:energy-usage} compares the hourly energy consumption profiles under the baseline and solar-integrated scenarios.
The baseline case shows higher overall grid energy consumption.
The solar-integrated scenario shows a significant shift in energy sourcing.

The economic performance is shown in Table~\ref{tab:sim_results_table}.
The integration of solar energy leads to a 3.7\% increase in overall profit.
This improvement is the result of 49.4\% savings in grid energy costs.
Additionally, the system achieved a 25.9\% decrease in machine startup costs, highlighting the benefits of optimizing production schedules around solar availability.
Incorporating solar energy results in a 5.6\% increase in buffer holding costs, suggesting that greater buffer flexibility is required to accommodate variations in solar energy availability.
Notably, the economic analysis does not include the initial investment costs associated with the solar energy infrastructure (rooftop PV, batteries, transmission lines, etc.).
Previous research reported that the average return on investment for the comparable scale of solar installations is approximately 24\%~\citep{formica2017return}.
Hence, the potential long-term economic benefits are significant.
If combined with U.S. tax incentives for solar PV systems, such as advanced energy project credit~\citep{IRS}, the investment presents an even more attractive long-run proposition.

\begin{table}[t]
\centering
\caption{Performance Evaluation}
\label{tab:sim_results_table}
\begin{tabular*}{\columnwidth}{@{\extracolsep{\fill}}lccc@{}}
\toprule
\textbf{Metric} & \textbf{Without} & \textbf{With} & \textbf{Diff} \\
& \textbf{Solar} & \textbf{Solar} & \textbf{\%}\\
\midrule
Profit (USD) & 1,5937 & 1,6530 & 3.7 \\
Grid cost (USD) & 1299 & 657 & -49.4 \\

Holding cost (USD) & 4926 & 5200 & 5.6 \\

Startup cost (USD) & 113 & 84 & -25.9 \\

Avg. price (USD/unit) & 82.50 & 74.77 & -9.4 \\

Production (unit) & 270 & 301 & 11.5 \\

Renewable (\%) & 0 & 51.5 & n/a \\
\bottomrule
\end{tabular*}
\end{table}

\section{Conclusion}
\label{sec:conclusion}

This paper proposes a bi-level MPC framework that jointly optimizes product prices and production scheduling while integrating renewable energy into manufacturing production.
The higher level sets prices to maximize profit and shape the market product demand in response to renewable energy usage based on price elasticity.
The lower level schedules production in runtime to meet demand and explicitly considers real-time electricity pricing and usage of onsite renewable sources.
Simulation results demonstrate that the proposed framework effectively balances profitability, demand fulfillment, and energy efficiency through reduced energy costs.
Future work will address stochastic demand, renewable availability, and distributed optimization for large-scale systems.

\bibliography{ifacconf}             
                                                   
                                                                         

\end{document}